\documentclass{aa}
\usepackage{graphicx}
\usepackage{txfonts}

\def\gtsim{ \lower .75ex \hbox{$\sim$} \llap{\raise .27ex \hbox{$>$}} } 
\def\ltsim{ \lower .75ex\hbox{$\sim$} \llap{\raise .27ex \hbox{$<$}} } 

\begin{document}
   
   \title{The cyclo--synchrotron process and particle heating through 
         the absorption of photons}
      
   \author{Katarzy\'nski K. \inst{1,2}, 
           Ghisellini G.    \inst{1},
           Svensson R.      \inst{3},
           Gracia J.        \inst{4}
          }

   \offprints{Krzysztof Katarzy\'nski \\kat@astro.uni.torun.pl}

   \institute{Osservatorio Astronomico di Brera, via Bianchi 46, Merate and via Brera 28, Milano, Italy  \and 
              Toru\'n Centre for Astronomy, Nicolaus Copernicus University, ul. Gagarina 11, PL-87100 Toru\'n, Poland \and
              Deceased \and
              IASA, Dept. of Physics, Univ. of Athens, Panepistimiopolis, 15784 Zografos, Athens 
             }

   \date{Received 12 October 2005 / Accepted 11 January 2006}

   \abstract{}
            {We propose a new approximation for the cyclo--synchrotron emissivity
             of a single electron. In the second part of this work, we discuss a 
             simple application for our approximation, and investigate the heating 
             of electrons through the self--absorption process. Finally, we 
             investigate the self--absorbed part of the spectrum produced by a 
             power--law population of electrons.
            } 
            {In comparison to earlier approximations, our
             formula provides a few significant advantages. Integration of the
             emissivity over the whole frequency range, starting from the proper
             minimal emitting frequency, gives the correct cooling rate for any
             energy particle. Further, the spectrum of the emission is well
             approximated over the whole frequency range, even for relatively low
             particle energies ($\beta \ll 0.1$), where most of the power is
             emitted in the first harmonic. In order to test our continuous
             approximation, we compare it with a recently derived approximation 
             of the first ten harmonics. 
             Finally, our formula connects relatively smooth to the synchrotron 
             emission at $\beta=0.9$.
            }
            {We show that the self--absorption is a very efficient heating
             mechanism for low energy particles, independent of the shape of the
             particle distribution responsible for the self--absorbed synchrotron
             emission. We find that the energy gains for low energy particles 
             are always higher than energy losses by cyclo--synchrotron emission. We 
             show also that the spectral index of the self--absorbed part of the
             spectrum at very low frequencies differs significantly from the well 
             known standard relation $I(\nu) \sim \nu^{5/2}$.
            }
            {}
             
\titlerunning{Cyclo--synchrotron emission}
\authorrunning{1st author et al.}   
\maketitle

\section{Introduction}

Synchrotron emission is well understood (see e.g. reviews 
by Ginzburg \& Syrovatskii 1965, 1969; Pacholczyk 1970)
and is thought to be responsible for a significant part of the 
radiation we receive from a variety of cosmic objects, such as 
supernova remnants, radio jets, compact radio sources, active 
galactic nuclei, and gamma--ray bursts.
The reverse process, synchrotron absorption, has only recently 
disclosed some novel features, when the attention of researchers 
shifted from what is the amount of absorbed flux of photons to what 
happens to the absorbing electrons.
In fact, they can absorb the {\it energy} of the photons and  thereby
change their initial distribution (Ghisellini, Guilbert \& Svensson 1988, 
hereafter GGS88), and/or absorb the {\it momentum} of the photons, with 
the possibility of bulk motion acceleration (Ghisellini et al. 
\cite{Ghisellini90}).

These authors demonstrate that the only stable equilibrium
solution of particles emitting and absorbing synchrotron radiation
is a relativistic or quasi--relativistic Maxwellian distribution.
This paper ended a long debate about the existence of another 
equilibrium solution: a power law of slope 3, i.e.
$N(\gamma)\propto \gamma^{-3}$, which was the main result
of the so-called ``Plasma Turbulent Reactor" (PTR)
theory, as introduced in a series of papers in the 1970s (Norman \cite{Norman77}, 
Norman and ter Haar \cite{NH75}, Kaplan \& Tsytovich \cite{KT73}).
Note also that the stability of this $N(\gamma)\propto \gamma^{-3}$
solution was already questioned by Rees (1967), stating that this
power law solution would evolve away from $\gamma^{-3}$, if slightly 
perturbed (see also the numerical results by McCray 1969, demonstrating
this instability).

One of the aims of this paper is to explicitly demonstrate, that the
$\gamma^{-3}$ distribution is not only unstable, but is 
not even an equilibrium solution.
To do so in an accurate way, however, it is necessary to
also consider the trans--relativistic regime, namely, the
cyclo--synchrotron emissivity and absorption coefficient.
This, of course, is the more complex regime, because the emitted power
is not concentrated at all in the first harmonic at the
typical synchrotron frequency (i.e. $\sim \gamma^2\nu_B$, where
$\nu_B$ is the Larmor frequency).
Recently, a useful approximation has been proposed by
Marcowith \& Malzac (2003), introducing polynomial expressions
for the first 10 harmonics for a range of particle energies.
They compare their results with an
existing {\it analytical} formula that tried to approximate
the emission (and the absorption) with smooth functions 
(i.e. not as sums of harmonics), as proposed by Ghisellini, Haardt, 
\& Svensson (1998). From this comparison it appears that there might 
still be room for improvement in this smooth, approximated function, 
which is the second main aim of our paper here.


We present our new approximation in Sect. 2, and compare it
with the Marcowith \& Malzac (2003) results.
We show that our approximation works well for particles with
$\beta<0.9$, where $\beta c $ is the particle velocity,
corresponding to $\gamma<2.3$.
For slightly higher energies, the standard synchrotron
formulae describe the shape of the emission well, but the
frequency--integrated emissivity must still be
corrected to become equal to the cooling rate (which is known
exactly for any particle energy).
Since we are interested in the {\it total} amount
of energy absorbed and lost by a single particle, in Sect. 3 we 
introduce a correction to the standard synchrotron formula, which is
important for $\gamma\gtsim 2.3$, but automatically negligible
for ultra-relativistic energies.

We then consider the rate of energy gains and losses
suffered by an electron of a given energy as a consequence
of cyclo--synchrotron emission and absorption (Sect. 4),
showing that only particles at a single energy can be in
equilibrium, where gains equal losses,
independently of the slope of the particle
distribution that produces the cyclo--synchrotron intensity.
Studying the low frequency part of the synchrotron
intensity in detail, we show that there are novel features 
below the Larmor frequencies that appear to have been 
overlooked in the past. These are presented in Sect. 5.
Finally, we draw our conclusions in Sect. 6.

\section{Approximation of the cyclo--synchrotron power spectrum}

The single particle cyclo-synchrotron power spectrum can be
approximated relatively well by a simple analytical formula. One of the
best approximations was proposed by Ghisellini, Haardt, \& Svensson
(\cite{Ghisellini98})
\begin{eqnarray}
P_c(\nu, p) & = & \frac{4}{3} \frac{\sigma_{\rm T} c U_B}{\nu_B} p^2
                f(p) \exp\left[f(p) 
                \left(1 - \frac{\nu}{\nu_B} \right)\right], \\
      f(p)& = & \frac{2}{1+ap^2}       
      \label{equ_fold}     
\end{eqnarray}
where $a=3$, $\nu_{\rm B} = e B / (2 \pi m_{\rm e} c)$ is the Larmor frequency, 
$U_B = B^2/(8\pi)$ the magnetic field energy density, $B$ the 
magnetic field strength, $p = \gamma \beta =\sqrt{\gamma^2 - 1}$ the 
dimensionless particle momentum, $\gamma$ the particle Lorentz factor 
related to the total energy by $E=\gamma m_e c^2$, $\beta$ the particle
velocity in units of $c$, and $\sigma_{\rm T},~m_e,~c$ are 
constant Thomson cross--section, electron rest mass, and the velocity of 
light, respectively. 
Eq 1 describes the power, integrated over the emission angles, of
an "average" electron: i.e. the power has been averaged over the pitch
angle, which is assumed to be distributed isotropically.
In this case the emissivity of a single electron is equal to the
emitted power divided by the solid angle factor $4\pi$. 
We use the term ``power spectrum" to indicate the 
emitted power as a function of frequency of an ``average" electron,
in the sense specified above.
This phenomenological formula has three advantages:
\begin{itemize}
\item[+]{can be integrated easily over the frequency range}
\item[+]{integration from $\nu=\nu_B$ up to infinity gives the correct 
         cooling rate
         \begin{equation}
          \dot{\gamma_c} m_e c^2 = \frac{4}{3} \sigma_{\rm T} c U_B p^2,
          \label{equ_coolratio}
         \end{equation}
        }
\item[+]{has the correct frequency dependence [$\exp(-\nu/\nu_B)$] at
         large harmonics ($\nu \gg \nu_B$)}
\end{itemize}
On the other hand, the formula also has three significant disadvantages:
\begin{itemize}
\item[--]{gives the correct cooling rate only if integrated from
  $\nu=\nu_B$,
  therefore cannot correctly describe
  the emission level below the frequency $\nu_B$}
\item[--]{does not approximate the emission spectrum well for
  $\beta < 0.5$, as we will show}
\item[--]{does not join smoothly to the synchrotron power spectrum.}          
\end{itemize}
Therefore, in order to improve this formula, we introduce three
important modifications.

First, the problem with the lower integration limit, ($\nu=\nu_B$)
 is solved by multiplying the formula with the term
\begin{equation}
g(\nu, p) = \frac{\nu-\nu_{\rm min}(p)}{\nu},
\end{equation}
where $\nu_{\rm min}$ indicates the minimal emitting frequency of the 
first harmonic
\begin{equation}
\nu_{\rm min}(p) = \frac{\nu_B}{\gamma  (1 + \beta)}.
\label{equ_numin}
\end{equation}
This modification significantly reduces the power emitted below 
the frequency $\nu_B$ in comparison to the original formula and provides
an automatic cut--off at the limiting frequency ($\nu_{\rm min}$). Note that
this additional term becomes unity for $\nu \gg \nu_{\rm min}$,
thus maintaining the original formula in this regime.
However, we introduce
a $\nu$--dependent term in front of the exponential
function ($\exp[f(1-\nu/\nu_B)]$), which also depends on the
frequency. Therefore, the integral
over the frequency range of the new formula must be expressed by the exponential
integral.

Second, to improve the shape of the spectrum for $\beta < 0.5$, we
replace the term $f(p)$ by a modified expression $f'(p)$ reading as
\begin{equation}
f'(p) = \frac{2}{1+ap^2} \frac{p^2+b}{p^2},
\end{equation}
where $a=3.65$ and $b=0.02$. The term $(p^2+b)/p^2$, which makes the difference 
between $f(p)$ and $f'(p)$, becomes equal to unity for $p \gg b$, and therefore
our approximation becomes equivalent to the original formula in this regime. The necessary
changes to the constant $a=3\to3.65$ is discussed in the next section.

Third, the new expression is normalized in order 
to yield the correct cooling rate (Eq. \ref{equ_coolratio}) when integrating 
over the frequency range. This normalization 
is done by multiplying with a factor 
\begin{eqnarray}
c(p) & = & \left\{ \exp \left[ f'(p) \left( 1-\frac{\nu_{\rm min}}{\nu_B} 
           \right) \right] \right. \nonumber \\
     & - & \left. f'(p) \frac{\nu_{\rm min}}{\nu_B} \exp[f'(p)] 
           {\rm Ei}_1\left[f'(p) \frac{\nu_{\rm min}}{\nu_B} \right] \right\}^{-1},
           \label{equ_fnew}
\end{eqnarray}
where ${\rm Ei}$ is the exponential integral (e.g. Press et al. \cite{Press89}).

Finally, the improved approximation for the cyclo-synchrotron power spectrum of
a single particle is given by
\begin{equation}
P_c(\nu, p) = \frac{4}{3} \frac{\sigma_{\rm T} c U_B}{\nu_B} p^2
            c(p) g(\nu, p) f'(p) \exp\left[f'(p) \left(1 - \frac{\nu}{\nu_B} 
            \right)\right]
            \label{equ_synemis}
\end{equation}
\begin{figure}[!p]
\resizebox{\hsize}{!}{\includegraphics{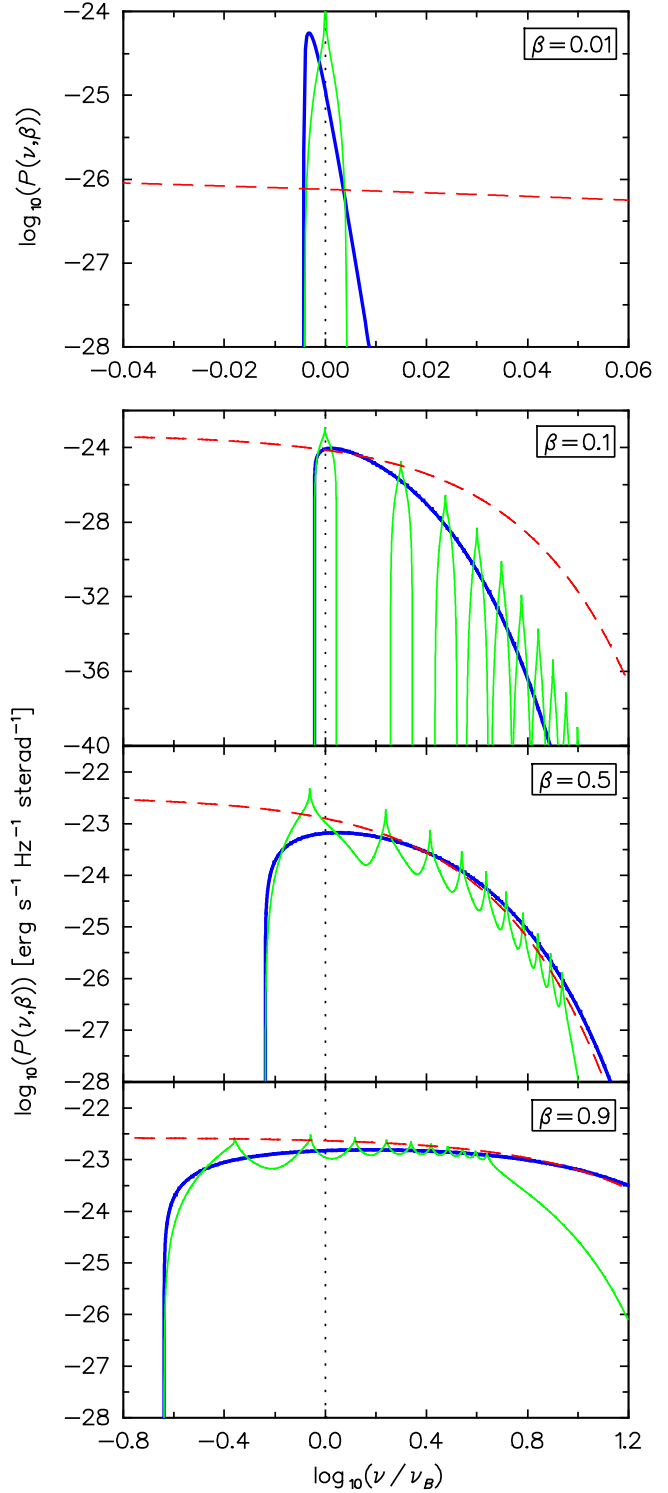}}
\caption{The comparison between different approximations of the cyclo--synchrotron 
         power spectrum of a single particle at four different energies. The thin 
         solid line in each panel shows the polynomial approximation of the first ten 
         harmonics provided by Marcowith \& Malzac (\cite{Marcowith03}). 
         The continuous approximation derived in this work is shown by the bold 
         solid line. The simple approximation proposed by Ghisellini et al. 
         (\cite{Ghisellini98}) is shown by the thin dashed line. Note that the 
         last approximation gives the correct cooling rate only if integrated 
         from frequency $\nu_B$ (indicated by the vertical dotted line).
        }
\label{fig_harmcomp}        
\end{figure}
In comparison to the original relation this formula has three significant
advantages:
\begin{itemize}
\item[+]{integrated from $\nu=\nu_{\rm min}$ gives the correct cooling rate
         for any energy particle}
\item[+]{describes the emission spectrum well, starting from the 
         minimal frequency for a wide energy range ($\beta=0.01\to 0.9$)}
\item[+]{provides a relatively smooth connection to the synchrotron power 
         spectrum at $\beta=0.9.$}
\end{itemize}
The price we pay for these improvements is that the integral 
over frequency of the new formula must be expressed in terms of an 
exponential integral. Moreover, in some sense our formula
averages over the harmonics, providing a continuous emission spectrum.
Therefore, it cannot be 
used for modeling the emission where the cyclo--synchrotron lines are 
observed directly (e.g. Pottschmidt et al. \cite{Pottschmidt05}). On the other 
hand, our approximation may have a wide range of applications in sources where 
the particle energy extends at least over one order of magnitude. In such 
a case the emission by particles with different energies may produce a
continuous spectrum, where the spectral lines are barely visible or completely 
negligible. 

The approximation of the cyclo--synchrotron emission provided 
by Marcowith \& Malzac (\cite{Marcowith03}) has been compared with the 
results of the precise numerical computations showing discrepancies 
that are less than 20\%.  
Of course, the discrepancy between our continuous approximation and the precise
calculation of (discrete) harmonics can be very large, if 
we compare our formula with the emission level between two 
well--separated harmonics.  
On the other hand, our main goal is to derive a formula that 
always provides the correct value of the total emitted energy.  
This is achieved through 
the normalization term (Eq. \ref{equ_fnew}) that independent of the 
values of the constants $a$ \& $b$, always provides the correct cooling ratio. 
This construction of the formula introduces a freedom in manipulating
of the spectral shape. Therefore, by choosing appropriate values 
for the parameters $a$ and $b$ 
we can approximate the spectra at different energies rather well.

In Fig. \ref{fig_harmcomp} we compare our new formula with the old 
relation and the approximation of the first ten harmonics provided by
Marcowith \& Malzac (\cite{Marcowith03}). A few general conclusions
can be drawn from this comparison:

\begin{itemize}
\item[$\bullet$]{
For very low particle energies ($\beta \ll 0.1$), most of 
the power is emitted in the first harmonic, i.e. in a very narrow frequency 
range. For this energy range, our approximation shows the biggest disagreement in 
comparison to the approximation of the first harmonic. However, in compared to 
the old formula, our approach gives significantly better results. 
}
\item[$\bullet$]{
In the intermediate energy range ($0.1 \lesssim \beta \lesssim 0.5$) the contribution of 
high--order harmonics to the total emission becomes more important. Moreover, with 
increasing particle energy, the emission from each harmonic spreads 
over a wider frequency range. The spectrum transforms from a set of discrete
lines into a continuous emission. Therefore, our approximation as a continuous 
function becomes more and more accurate with increasing particle energy.
}
\item[$\bullet$]{
Finally, in the range of relatively high energy ($0.5 \lesssim \beta \lesssim 0.9$), 
the emission is dominated by the high--order harmonics. Our approximation
gives the best agreement with the approximation for the first ten harmonics. The 
old formula also provides a relatively good approximation in this energy range. However, 
the total radiated power calculated from this formula is correct only if  
integrated starting from frequency $\nu_B$. 
Therefore, it cannot describe the spectrum below $\nu_B$, 
since it overestimates the total emitted energy if integrated from $\nu_{\rm min}$
The approximation of the first ten harmonics does not  
provide a correct value of the total emitted energy, either
due to the fact that the ten harmonics considered  
provide a good description only up to 
$\nu \sim 4\nu_B$, while there is still considerable power above, too.
For this particular particle energy range, many more harmonics are needed 
to provide the correct value of the total emitted energy.
}
\end{itemize}
Above $\beta=0.9$, the synchrotron power spectrum can be used; however, this 
coefficient also requires some correction, which we describe in the next 
section.

\section{Correction of the synchrotron power spectrum}

The synchrotron power spectrum from a single particle in a random magnetic field, 
integrated over an isotropic distribution of pitch angles, has been derived
by Crusius \& Schlickeiser (\cite{Crusius86}) and GGS88:
\begin{eqnarray}
P_s(\nu, \gamma) & = & \frac{3 \sqrt{3}}{\pi} \frac{\sigma_{\rm T} c U_B}{\nu_B} ~ x^2 
                \times \nonumber \\
                 & \times & \left\{ K_{4/3}(x) K_{1/3}(x) - 
                   \frac{3}{5} x \left[ K^2_{4/3}(x)-K^2_{1/3}(x) \right] \right\},
\end{eqnarray}
where $x=\nu/(3 \gamma^2 \nu_B)$ and $K_y(x)$ is the modified Bessel function of 
order $y$. 
This formula does not provide the correct cooling rate for 
$\gamma \lesssim 15$, where it overestimates the total emitted energy. 
Since the spectral shape, however, is approximately 
correct even for low particle energies ($\gamma<15$),
we multiply the original formula by a simple correction 
term that only depends on particle energy
\begin{equation}
s(\gamma) = \frac{\dot{\gamma_c}(\gamma)}{\int^{\infty}_{\nu_{\rm min}} 
            P_s(\nu, \gamma) d \nu}.
\end{equation}
\begin{figure}[!t]
\resizebox{\hsize}{!}{\includegraphics{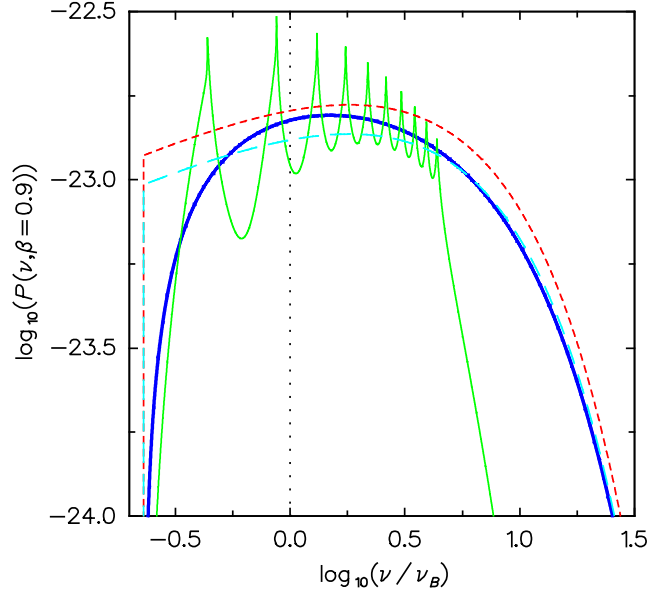}}
\caption{The approximated cyclo--synchrotron power spectrum and the corrected 
         synchrotron spectrum for $\beta=0.9$. The thin solid line shows the
         approximation of the first ten harmonics provided by Marcowith \& Malzac 
         (\cite{Marcowith03}), the continuous approximation derived in this
         paper is shown by the bold line, the synchrotron power spectrum derived
         by Crusius \& Schlickeiser (\cite{Crusius86}) and GGS88 is shown by the short
         dashed line and the long dashed line shows the corrected synchrotron
         spectrum.
        }
\label{fig_syncomp}        
\end{figure}
The difference between the cooling rate ($\dot{\gamma_c}$) and the
frequency integrated synchrotron power disappears
for $\gamma\gg 15$; therefore, the correction term reduces to unity
for high particle energies.

In Fig. \ref{fig_syncomp} we compare our approximated cyclo--synchrotron 
power spectrum with the corrected synchrotron spectrum at $\beta=0.9$. Since
the correction term for the synchrotron formula only depends on the
particle energy, the correction only affects the normalization of the
synchrotron spectrum. The figure also shows, that our approximation joins
relatively smoothly to the corrected synchrotron spectrum at $\nu\gg\nu_B$.
However, in order to achieve this smooth connection, we had to modify the constant $a$
(Eqs \ref{equ_fold} and \ref{equ_fnew}), that controls the spectrum shape at
$\nu\gg\nu_B$.  
The constant $b$ in our formula 
controls the spectral shape for $\beta\lesssim 0.5$. Note that
any modification of the parameters $a$ or $b$ changes the spectral shape and
thus, in principle, also the total emitted energy. This problem has
been solved through the normalization term (Eq. \ref{equ_fnew}), which also contains 
the parameters $a$ \& $b$, and thus changes the level of the spectrum in 
order to keep the correct value of the total emission.

\section{Particle heating through the absorption of photons}

We present a simple application for our approximation of the cyclo--synchrotron 
power spectrum and the corrected synchrotron emission coefficient. 
We analyze the amount of energy gain
corresponding to the cyclo--synchrotron absorption process.
This process may lead to an efficient exchange 
of the energy, and may therefore provide a very powerful heating mechanism
for low energy particles. This kind of heating is a stochastic process. 
This process of competing radiative cooling and radiative heating, through emission 
and re-absorption, respectively, leads to the
accumulation of most particles around the equilibrium energy ($\gamma_e$), 
where heating and cooling are in balance. 
In an ideal case, such a competition would transform any initial 
particle distribution into a thermal Maxwellian distribution with its maximum at the 
equilibrium energy.

In our test, we investigated the self--absorption of the radiation field 
produced by electrons with a power law energy distribution 
\begin{equation}
N(\gamma) \propto \gamma^{-n},
\end{equation}
located inside a homogeneous spherical volume that is filled by a tangled magnetic 
field. We compare the efficiency of the heating and the cooling processes and discuss 
the possible values of the equilibrium energy for different slopes of the electron 
energy distribution.

The emission coefficient for any electron energy distribution
$N(\gamma)$ is defined by
\begin{equation}
j(\nu) = \frac{1}{4 \pi} \int N(\gamma) P(\nu, \gamma) d \gamma,
\end{equation}
where $P(\nu, \gamma)$ is either the cyclo--synchrotron or the corrected synchrotron 
power, depending on the value of $\gamma$. The self--absorption coefficient for any
particle distribution is given by Le Roux (\cite{LeRoux61})
\begin{equation}
k(\nu) = \frac{1}{8 \pi m_e \nu^2} \int \frac{N(\gamma)}{\gamma p} 
                 \frac{d}{d\gamma}\left[ \gamma p P(\nu, \gamma) \right] 
                 d \gamma.
\label{equ_abscoeff}                 
\end{equation}

For the calculations we assume a constant intensity of the radiation field 
inside the source, which is approximated by
\begin{equation}
I(\nu) = C \frac{j(\nu)}{k(\nu)} 
           \left\{1 - \exp[-k(\nu) R] \right\},
\end{equation}
where $R$ is the source radius. The value of the parameter $C$ depends in 
principle on the position within the source. In the center of our homogeneous source 
$C=1$ should be used, whereas on the source surface we should
apply $C=1/2$ (Gould \cite{Gould79}). However, for the sake of simplicity
we assume a constant value of the intensity for the whole absorbing
region, and use an average value of $C=3/4$. 

Finally, the absorption or heating efficiency
for any radiation field is described by (Ghisellini \& Svensson
\cite{Ghisellini91}) as
\begin{equation}
\dot{\gamma_a}(\gamma) = \frac{1}{m_e c^2} \frac{1}{\gamma p} \frac{d}{d \gamma}
\left[\gamma p \int_0^{\infty} \frac{I(\nu)}{2 m_e \nu^2} P(\nu, \gamma) d \nu
\right].
\end{equation}
We can easily estimate the absorption efficiency in an asymptotic case where 
the particle energy is characterized by $\gamma \to 1$ ($\beta \ll 0.1$). In 
this case, most of the energy is emitted in the first 
harmonic, i.e. in a very narrow frequency range (see Fig.~\ref{fig_harmcomp} for
$\beta=0.01$). Therefore, the cyclo--synchrotron power spectrum can be 
approximated by a $\delta$--function
\begin{equation}
P_c(\nu, \gamma \to 1) = m_e c^2~\dot{\gamma}_c(p)~\delta(\nu-\nu_B)
\end{equation}
which results in a constant value of the absorption efficiency
\begin{equation}
\dot{\gamma_a}(\gamma \to 1) = \frac{4}{3} c \sigma_{\rm T} U_B
                               \frac{I(\nu_B)}{2 m_e \nu_B^2}
                               \left( 3\gamma + \frac{p^2}{\gamma} \right)
\to {\rm constant}.
\label{equ_delta}
\end{equation}

Comparing this result with the cooling efficiency, which is always proportional
to $p^2$ (Eq. \ref{equ_coolratio}), we see that
{\it for low energy particles, 
heating due to the self--absorption will always overcome radiative 
losses.} 
In other words, for a power law particle energy distribution,
the heating term is proportional to $p^2$ for high values of $p$
(in the self--absorbed regime);
on the other hand, in the low energy limit $p\to 0$ (or $\gamma\to 1$),
 the heating term {\it must} always be constant.
This implies, that there always is only one specific energy value for which
the heating and cooling terms are equal.
To illustrate this point, let us compare this result 
with the equilibrium solution [$N(\gamma)\propto \gamma^{-3}$]
proposed in the ``turbulent reactor" scenario.
In this case, the (analytical) solution is obtained by
considering only the relativistic regime and further assuming  an
infinite source (i.e. infinite self--absorption frequency).
In other words, the radiation intensity is assumed to be proportional
to $\nu^{5/2}$ for all frequencies, with a normalization that
depends on the slope of the electron distribution (see also
Rees \cite{Rees67} and Mc Cray \cite{McCray69}).
If one assumes that the particle distribution is a power law, but 
truncated at some low energy $\gamma_{\rm min}$ (to self consistently 
use the emissivity and absorption processes in the relativistic regime)
then the radiation field is not $\propto \nu^{5/2}$ in the entire
frequency range, but only above $\sim \gamma_{\rm min}^2\nu_B$.
Below this frequency, $I(\nu) \propto \nu^2$ (e.g. Rybicki \& 
Lightman \cite{Rybicki79}),
making the electrons gain energy (through absorption) 
at a slightly higher rate than what is found by assuming 
$I(\nu)\propto \nu^{5/2}$.
If instead one assumes a particle distribution extending
towards mildly relativistic energies, then one obtains that
$\dot\gamma_a$, for small particle energies
becomes constant and thus clearly larger than 
the extrapolation of the $\dot\gamma_a \propto p^2$ law.
Concerning the other energy extreme, a source
of finite size becomes transparent at some finite
value of the self--absorption frequency $\nu_t$.
Therefore the radiation field inside the source 
for frequencies close to and above $\nu_t$ is
no longer proportional to $\nu^{5/2}$, with a
corresponding decrease in the energy gain rate
for high--energy particles.
These are the reasons a particle distribution
$N(\gamma)\propto \gamma^{-3}$ is not an equilibrium 
solution.
For any given value of the slope of the particle distribution,
equilibrium is always achieved at only one specific energy.

\begin{figure*}[!t]
\resizebox{\hsize}{!}{\includegraphics{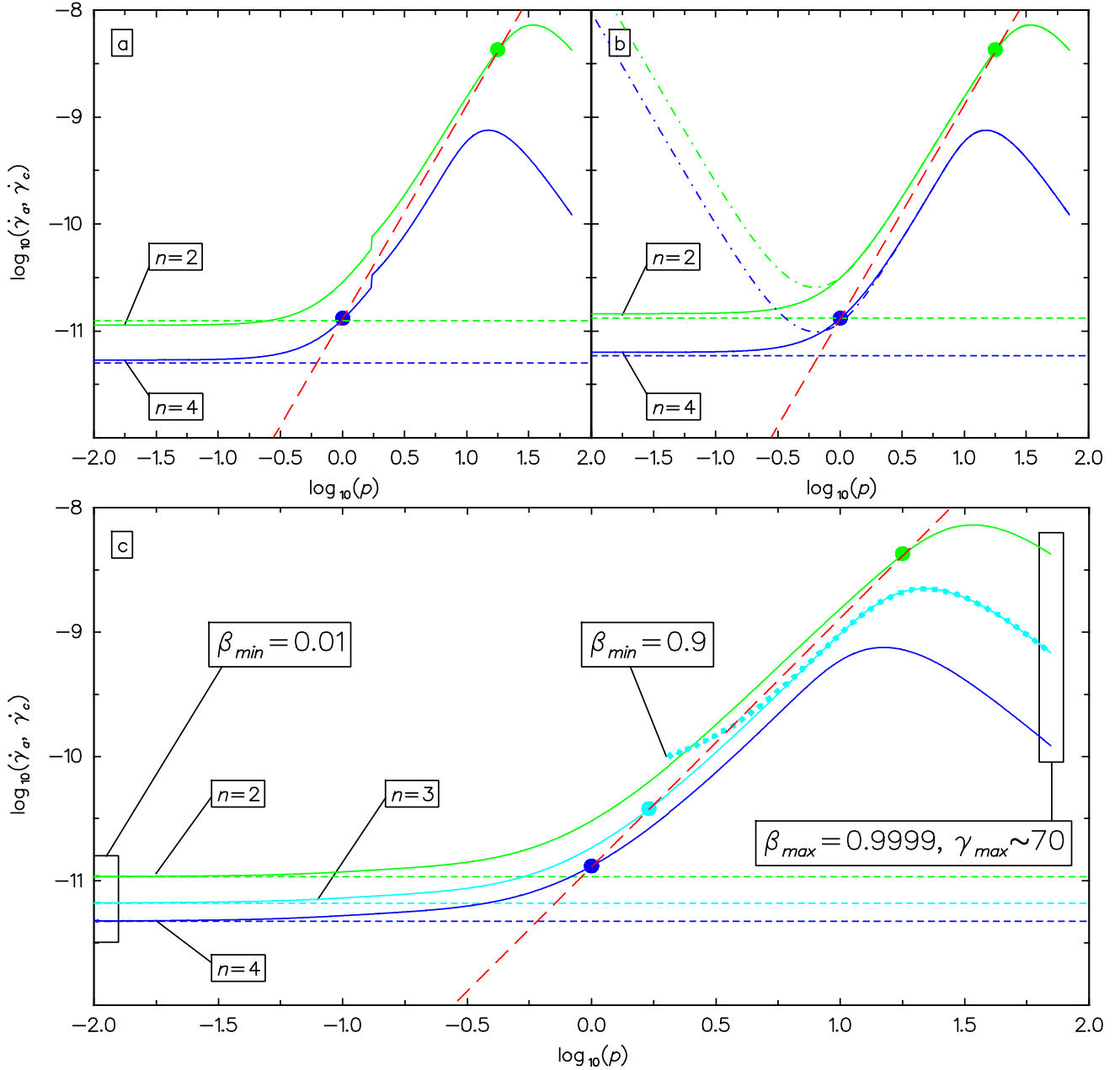}}
\caption{The efficiency of absorption (solid lines) versus the cooling 
         efficiency (long dashed lines) for the power law particle energy 
         distribution ($\beta_{\rm min} = 0.1 \to \beta_{\rm max} = 0.9999$, 
         $\gamma \simeq 70$) for $n=2$, $n=4$, and $n=3$ (the last only in the 
         main panel - c). The left upper panel (a)
         shows the absorption efficiency calculated from the approximation of
         the cyclo--synchrotron emission derived by Ghisellini et al. 
         (\cite{Ghisellini98}) and the standard synchrotron emission coefficient 
         provided by Crusius \& Schlickeiser (\cite{Crusius86}) and GGS98. In the 
         upper right panel (b), we show the absorption efficiency calculated only from 
         the corrected synchrotron emission. The dash--dot lines on this panel 
         indicate the result obtained from uncorrected synchrotron emission.
         In the main panel (c). the new cyclo--synchrotron approximation and the 
         corrected synchrotron emission were used. The short dashed lines in each 
         panel indicate a constant value of the absorption efficiency estimated 
         for $\gamma \to 1$ (Eq. \ref{equ_delta}). The equilibrium energies are 
         indicated by dots. Note that in the main panel we also show the absorption 
         efficiency calculated for $\beta_{\rm min} = 0.9$ and $n=3$ (bold dotted 
         line).
        }
\label{fig_abseff}        
\end{figure*}

In Fig. \ref{fig_abseff} we compare the absorption efficiency with 
the cooling rate of a power law particle energy distribution from 
$\beta_{\rm min}=0.01$ to $\beta_{\rm max}=0.9999$ ($\gamma_{\rm max} 
\simeq 70$) with three different slopes ($n=2$, $n=3$, and $n=4$).
We performed the computations using 
three different approaches for the calculation of the cyclo--synchrotron 
power spectrum in order to test the formulae derived in this work. 

First, we used the old formula for the cyclo--synchrotron power spectrum 
(Eq. \ref{equ_fold}) and the standard, uncorrected synchrotron emission 
coefficient (Eq.~\ref{equ_synemis}). The result of these computations is 
presented in the upper left panel of Fig. \ref{fig_abseff}. In this particular 
case, the transition from the old cyclo--synchrotron power spectrum to the 
uncorrected synchrotron emission at $\gamma=2$, produces a clearly visible 
discontinuity in the absorption efficiency. The discontinuity is related 
to the fact, that the uncorrected synchrotron emission overestimates 
the total emitted energy for $\gamma \lesssim 15$. Moreover, for low 
energy particles ($p<0.1$) the level of the absorption efficiency does not 
agree with the constant level estimated from the $\delta$--approximation 
(Eq. \ref{equ_delta}), which is indicated by the horizontal lines. 

In the second test, presented on the main panel of Fig.~\ref{fig_abseff}, 
we used our new approximation of the cyclo--synchrotron power spectrum and the 
corrected synchrotron emission coefficient. Since both expressions provide the 
correct value for the total emitted energy, the transition
from one spectrum to the other (this time at $\beta=0.9$) does not 
produce any discontinuity in the absorption efficiency. Moreover, with our new
approximation of the cyclo--synchrotron emission, the absorption 
efficiency at low particle energies agrees well with the $\delta$--approximation.

Finally, we used only the corrected synchrotron
emission coefficient for the whole energy range (right panel in Fig.~\ref{fig_abseff}).
No discontinuity is present, since we used only one formula. However, this approach, 
as well as the old cyclo--synchrotron power spectrum in the first test, do not agree
very well with the $\delta$--approximation. Note that for low energy 
particles, the uncorrected synchrotron emission gives an absorption efficiency
that is a few orders of magnitude larger than the efficiency obtained from the other 
approximations.

These three approaches for the calculating of the absorption
efficiency, qualitatively give the same results, but differ
in the quantitative details, which indicate that the new approximation for 
the cyclo--synchrotron power spectrum together with the corrected synchrotron emission 
coefficient, provides the most precise description. 

Our tests indeed show that the absorption efficiency for the very low--energy
particles becomes independent of the particle energy and is significantly
higher than the cyclo--synchrotron cooling ratio. 
Therefore, self--absorption will 
always cause a strong heating of the low--energy particles. 
The equilibrium energy depends strongly on the slope of the particle 
spectrum. 
For $n=2$ the equilibrium energy is very close to the peak in the
absorption efficiency (the $\dot\gamma_a(p)$ curve), which
is related to the maximum in the self--absorbed
spectrum ($\nu_t$). 
If the particle spectrum is steeper ($n=4$), the equilibrium 
is taken at a lower energy, but close to the energy
of those particles emitting at the peak of the synchrotron spectrum.
The $n=3$ case is particularly interesting, since it corresponds
to the previously claimed equilibrium solution.
Contrary to this claim, heating and cooling also balance in this case only
at a specific energy.
This is, on one hand, due to the finite size of the source that limits
the range of possible momenta of electrons that emit and absorb radiation 
efficiently and on the other hand, more importantly, due to the trans-- and 
sub--relativistic regime where low energy particles always 
gain more energy than they loose.

The value of the equilibrium energy strongly depends on the minimum 
and maximum energy of the particles. In our tests the equilibrium 
energy for $n<3$ depends on the self-absorption frequency ($\nu_t$). 
However, for relatively low value of $\gamma_{\rm max}$ (e.g. 7 instead of 
70 in our particular calculations), the emission should be absorbed at
all frequencies of a completely optically thick source. In 
such a case, the equilibrium energy depends directly on $\gamma_{\rm max}$.
For $n \ge 3$ the equilibrium energy depends on the minimal energy
of the particles. In Fig. \ref{fig_abseff} we show the heating 
efficiency calculated for a relatively high value of the minimum 
particle energy $\beta_{\rm min} = 0.9$. The 
value of the equilibrium energy increases with the increasing
minimum energy.

Note that for sake of simplicity,  our tests assume
a stationary state, where the equilibrium energy is simply given by 
the equilibrium between the heating and cooling rates. In reality, the 
system evolves and the physical conditions inside the 
source change. 
The initial power law, or any other particle distribution, 
will be transformed into a thermal or quasi--thermal spectrum (see GGS88). 
Also, the equilibrium energy in such an evolving source may be different from 
the energy estimated from the simple stationary analysis. 
This does not change that there is only one preferred equilibrium 
energy around which most of the particles will be accumulated, forming thermal 
or quasi--thermal distribution. A complete description of this time 
dependent evolution will be the main focus of our future study.

\section{The self--absorbed spectrum}
\label{sec_rad}

The detailed analysis of the cyclo--synchrotron emission shows that the 
self--absorbed part of the spectrum generated by the electrons with a
power--law energy distribution, is slightly different from the well 
known power law relation $I(\nu) \propto \nu^{5/2}$. In this section
we analyze the emission of our homogeneous source, assuming different slopes
of the particle spectrum, and discuss the reasons for the deviations 
from the standard $\nu^{5/2}$ spectrum.

\begin{figure}[!t]
\resizebox{\hsize}{!}{\includegraphics{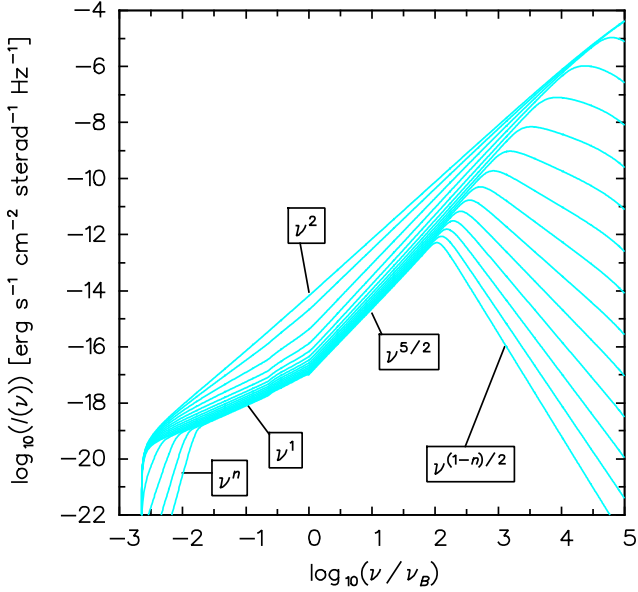}}
\caption{The self--absorbed part of the cyclo--synchrotron and the synchrotron 
         emission from a homogeneous spherical source. The intensities were 
         calculated for a range of different indices of the power law particle energy 
         spectrum, starting from $\gamma^{1}$ up to $\gamma^{-25/3}$ with a finite step size
         $\Delta n =2/3$. The spectrum on the top [$I(\nu) \propto \nu^2$ in 
         almost the 
         entire frequency range shown] corresponds to $n=-1$, whereas the 
         lowest spectrum was calculated for $n=25/3$. The limiting values of 
         the spectral indices presented in this 
         figure are discussed in the Sect. \ref{sec_rad}.
        }
\label{fig_rad}        
\end{figure}

The observed intensity of the emission from the homogeneous spherical 
source is given by
\begin{equation}
I(\nu) = \frac{j(\nu)}{k(\nu)}
         \left(1 - \frac{2}{\tau^2} 
         \left[1 - {\rm e}^{-\tau} (\tau+1)
         \right]
         \right)
\end{equation}
where $\tau = 2 R k$ is the optical depth (e.g. Bloom \& Marscher \cite{Bloom96}).
We calculate the intensity for a range of power law slopes, starting from $n=-1$ 
up to $n=25/3$ with finite steps of 
$\Delta n = 2/3$. Our results are shown in Fig. \ref{fig_rad}.
For $n\le -2/3$, the spectral index is constant at $\alpha=-2$, in almost the whole 
self--absorbed part of the spectrum. 
In the range $-2/3 \le n \le 5/3$, the slope
of the spectrum changes from $\alpha=-2$ to $\alpha=-1$ for $\nu<\nu_B$, and
from $\alpha=-2$ to $\alpha=-2.5$ for $\nu>\nu_B$. Above the limiting value 
$n=5/3$, the spectral index remains constant 
\begin{eqnarray}
\alpha=&-1  ~~~~~~  {\rm for}~~~\, \nu<\nu_B \nonumber, \\
\alpha=&-5/2~~~  {\rm for}~~~ \nu>\nu_B \nonumber.
\end{eqnarray}
Note that, for relatively steep particle distributions ($n \ge 17/3$ in this 
particular case) close to the low frequency cut--off, the spectrum may go over
into a power law with the index $\alpha=-n$. Therefore, in such a case the 
self--absorbed part of the spectrum is described by three different indices 
$\alpha=-n\to-1\to-5/2$. 
It is relatively easy to understand this  
specific evolution of the self--absorbed spectrum, if we analyze the limiting 
cases $n\le-2/3$ and $n\ge5/3$. 

In the first case ($n\le-2/3$), the emission in the whole frequency range is dominated 
by the synchrotron radiation produced by the highest energy electrons.
The synchrotron power spectrum of a single particle or 
monoenergetic population of the particles can be approximated below the 
peak frequency ($\nu_p(\gamma)$) by
\begin{equation}
P_s(\nu, \gamma) \propto \nu^{1/3}~\gamma^{-2/3}.
\label{equ_synemisapp}
\end{equation}
Integrating this formula over the power--law electron energy spectrum with the
index $n\le-2/3$, we obtain
\begin{equation}
j(\nu) \propto \nu^{1/3}~~~{\rm for}~~~\nu_{\rm min}(\gamma_{\rm max}) \ll 
   \nu \ll \nu_p(\gamma_{\rm max}).
\end{equation}
Calculating the absorption in the same way, Eq. \ref{equ_abscoeff}, we obtain
\begin{equation}
k(\nu) \propto \nu^{-5/3}~~~{\rm for}~~~\nu_{\rm min}(\gamma_{\rm max}) \ll 
  \nu \ll \nu_p(\gamma_{\rm max}),
\end{equation}
and this gives $I(\nu) = j(\nu)/k(\nu) \propto \nu^2$ in the self--absorbed part of the spectrum.

In the second limiting case ($n\ge 5/3$), the spectral index above $\nu_B$ is 
equivalent to the well--known solution ($\alpha=-5/2$), and we only discuss the 
reason for the flattening of the spectrum ($\alpha=-1$) below this frequency. 
Around $\nu_B$ the emission is dominated by the cyclo--synchrotron radiation
of the low--energy particles.
However, for $\nu \ll \nu_B$ the emission becomes dominated by the tail of the 
synchrotron emission of the high--energy particles. 
Therefore, we can again use the approximation of the single--particle power 
spectrum (Eq. \ref{equ_synemisapp}).
Integrating this approximation over the power law electron spectrum and 
neglecting the lower integration boundary, we obtain 
$j(\nu) \propto \nu^{1/3} \gamma^{1/3-n}_{\rm max}$. 
According to Eq. \ref{equ_numin} the maximum
energy is directly related to a given frequency $\nu \sim 1/ \gamma_{\rm max}$.
Therefore, we obtain
\begin{equation}
j(\nu) \propto \nu^{n}~~~{\rm for}~~~\nu_{\rm min}(\gamma_{\rm max}) \ll \nu \ll \nu_B.
\label{equ_emisidx}
\end{equation}
In the same way, we can integrate the absorption coefficient Eq.~\ref{equ_abscoeff}
obtaining $k \sim \nu^{-5/3} \gamma^{-n-2/3}_{\rm max}$, which gives
\begin{equation}
k(\nu) \propto  \nu^{n-1}~~~{\rm for}~~~\nu_{\rm min}(\gamma_{\rm max}) \ll \nu \ll \nu_B.
\end{equation}
Finally, this gives $I(\nu)= j(\nu)/k(\nu) \propto \nu^1$ in the self--absorbed part of the 
spectrum below the  frequency $\nu_B$.

The low--frequency emission ($\nu \ll \nu_B $) can be absorbed only by the 
high--energy particles. For relatively steep particle spectra, the density of these
particles can be too small to efficiently absorb the low--frequency radiation.
Therefore, the source may again become optically thin and the spectral index 
equivalent to the index of the emission coefficient $I(\nu) \propto \nu^n$
(Eq.~\ref{equ_emisidx}).

When calculating the spectra presented in Fig. \ref{fig_rad}, we used our
new formula for the cyclo--synchrotron power spectrum and the corrected 
synchrotron--emission coefficient. There is no smooth transition between these formulae at
$\beta=0.9$ for $\nu \lesssim \nu_B$ (see Fig. \ref{fig_syncomp}), but this is barely 
visible in our spectra. The effects discussed in this section depend mostly on 
the synchrotron emission; therefore, the corrected synchrotron emission can be 
used in whole energy range to get almost identical results.

The modifications of the self--absorbed spectrum that we discuss appear at
relatively low frequencies. 
Therefore, in most astrophysical objects such effects are not observable.
However, some effects might be visible in some specific physical conditions. 
One example might be the synchrotron radiation (in its self--absorbed
portion) produced by steep power-law distributions of particles 
in highly magnetized sources ($B \sim 10^{7\to10}$ [G]). 
Note that isotropic distribution of the pitch--angles, assumed in order 
to derive the emissivity formulae presented in this paper, may not always 
be valid, especially in very highly magnetized sources.

\section{Discussion}

We have derived a new approximation for the cyclo--synchrotron power spectrum of a 
single particle and compared it with the approximation to ten first harmonics of 
the cyclo--synchrotron emission provided recently by Marcowith \& Malzac 
(\cite{Marcowith03}).
In comparison to the other approaches, our approximation self--consistently 
provides the correct value of the total emitted energy over the whole range of 
the particle energies. Moreover, our approach 
describes the spectrum of the emission in the range $0.1<\beta<0.9$ relatively well.
Finally, the approximation provides a relatively smooth connection with the corrected
synchrotron emission at $\nu>\nu_B$ for $\beta=0.9$.

All these results are useful when one needs fast computational tools
to derive the cyclo--synchrotron emission and absorption, instead of
using the exact expressions, which require much more computing time.

The application we will pursue is to study in detail the evolution
of the emitting particle distribution subject to acceleration and/or injection
of new particles, radiative and Coulomb cooling, and heating due to the
synchrotron self--absorption. 
For the moment, we have instead analyzed a simpler process that, however, 
requires a careful treatment of the trans--relativistic regime.
We have demonstrated that a power--law  distribution of electrons
emitting and absorbing cyclo--synchrotron photons can {\it never}
be a steady solution.
Energy losses and gains {\it always} equal each other at a particular energy 
$\gamma_e$, and not over a range of energies.
This is contrary to previous claims that a distribution 
$N(\gamma)\propto \gamma^{-3}$ can be an equilibrium solution (e.g. Kaplan \& Tsytovich 1973).
The reason is that, at low enough, sub-- or trans--relativistic energies, 
energy gains always exceed losses and, at the other extreme, the absorption becomes
less efficient because the source becomes transparent (unless it is infinite
in size).
Particles therefore will tend to accumulate at $\gamma_e$, changing the shape 
of the energy distribution they initially belonged to.

For the steep particle spectrum ($n\geq 3$), the equilibrium energy ($\gamma_e$)
is small. This motivated our detailed study of the trans--relativistic 
cyclo--synchrotron regime.
However, we have shown that different approximations lead to very
similar results for the amount of energy gains experienced by
the particle.
This is due to the fact that in calculating this quantity we must consider
frequency--integrated expressions, with the consequent loss of details
concerning the shape of the power spectrum.
What matters is mainly that the frequency integrated spectrum equals
the correct cooling rate (i.e. $\propto p^2$).

Exchanging photons through emission and absorption allows particles
to exchange energy, independently of Coulomb collisions.
This is a very important thermalization process in those
magnetized, hot and rarefied plasma where Coulomb collisions
are rare.
Cyclo--synchrotron absorption transforms an initially non--thermal 
distribution into a Maxwellian in just a few cooling times (GGS88).

\begin{acknowledgements}
We thank the anonymous referee for a number of constructive 
comments that improved the paper. GG always remembers with deep gratitude 
the friendship with Roland Svensson who passed away prematurely. This work
is to partly acknowledge large number of ideas and work that he 
initiated, but did not succeed in publishing. We acknowledge the EC funding 
under contract HPRCN-CT-2002-00321 (ENIGMA network).
\end{acknowledgements}

\end{document}